\documentclass{aastex}          
\usepackage{amssymb}
\usepackage{graphicx,fleqn,color}
\usepackage{latexsym}
\usepackage{amsfonts}
\usepackage{amsmath}
\usepackage{amssymb}

\begin{document}

\title{On distortion of the background radiation spectrum by wormholes:
kinematic Sunyaev-Zel'dovich effect}
\author{A.A. Kirillov\altaffilmark{1} and E.P. Savelova\altaffilmark{1}}

\begin{abstract}
The problem of scattering of the background radiation on relic cosmological
wormholes is considered. It is shown that static wormholes do not perturb
the spectrum at all. The presence of peculiar velocities of wormholes
results in a distortion of the CMB spectrum which is analogous to the
kinematic Sunyaev-Zel'dovich effect. In the first order in $v/c$ the
distortion of CMB cannot be separated from from the Compton scattering on
electrons. In next orders the scattering on wormholes exhibits some
difference from the Compton scattering. High-energy cosmic-ray particles
spectrum does not change the form by KSZ, but undergoes a common Doppler
shift. Such features may give a new tool to detect the presence of relic
wormholes in our Universe.
\end{abstract}


\altaffiltext{1}{Bauman Moscow State Technical University, Moscow, 105005,
Russia}

\section{Introduction}

As it was recently shown some basic difficulties of cold dark matter models (%
$\Lambda $CDM) can be cured by the presence of relic cosmological wormholes %
\citep{KS11,ks16,KS17}. To avoid misunderstanding we point out that relic
wormholes are not going to replace completely the dark matter paradigm,
since there exist phenomena related to dark matter which wormholes unable to
explain. The existence of relic wormholes however is not in a conflict with
the simultaneous existence of dark matter particles, the so-called WIMPs
(weakly interacting massive particles). Save the dark matter phenomena
observed in astrophysics (dark matter halos in galaxies, CMB spectrum,
observed structures, etc.), the presence of WIMPs is well motivated by
numerous problems of the Standard Model in particle physics, e.g., see the
list in \citep{Feng10}. In particular, the observed high-energy cosmic-ray
electrons and positrons \citep{RS} may enable the observation of phenomena
such as dark-matter particle annihilation or decay \citep{grib}. Relic
wormholes do not produce such an effect, though spherically symmetric
wormholes collapse and form black holes and may produce all astrophysical
phenomena related to them.

We however point out that WIMPs may have the direct relation to
virtual wormholes. Such wormholes have virtual character and describe
quantum topology fluctuations \citep{S15,S16}. It was shown recently in %
\citep{KS15} that for all types of relativistic fields, the scattering on
virtual wormholes leads to the appearance of additional very heavy
particles, which play the role of auxiliary fields in the invariant scheme
of Pauli--Villars regularization. In the simplest picture the mas spectrum
of such additional particles starts from the Planck value $M_{pl}$
and is completely determined by parameters of the vacuum distribution of
virtual wormholes. It is important that such additional particles are
generated for all sorts of particles in the Standard Model and have a
discrete spectrum of the more increasing masses. For example, standard
massless photons are accompanied with massive photons with masses
$M_{i}=a_{i}M_{pl}$, where coefficients $a_{i}$ ($a_{1}<a_{2}<...$)
are expressed via the distribution of virtual wormholes \citep{KS15}.
On the very early stage of the development of the Universe such particles
were in an abundance in the primordial hot plasma. During the expansion the
Universe cools and most of such particles decay. At least all such particles
decay if they are involved in the strong or electromagnetic interactions.
However weakly interacting particles may survive till the present days and
they indeed may play the role of dark matter (e.g., extremely massive
gravitons, neutrinos, etc.).The decay of such superheavy particles
into unstable particles with large mass is described by \citep{grib}, while
their subsequent decay into quarks and leptons leads to events in cosmic
rays. In particular, the detected break in the teraelectronvolt cosmic-ray
spectrum of electrons and positrons \citep{RS} can be interpreted as the
trace of the decay of two sorts of such particles with different masses $
M_{1}\ll M_{2}$. The values $M_{i}$ determine the absolute
boundaries of the respective spectra, while the factor
$\exp \left( -\frac{\Delta M}{T_{c}}\right) $ (where $T_{c}$ is the
temperature at which the primordial content of such particles had been tempered)
determines the ratio of the respective number of events. One may expect that analogous
break should be observed and for higher energies as well.

In astrophysical picture relic wormholes produce also a number of effects
analogous to effects from dark matter particles and, therefore, the number
density of such particles in galactic halos may essentially change, when the
presence of relic wormholes is taken into account. Indeed, as it was
demonstrated by \citep{KS11} at very large scales wormholes contribute to
the matter density perturbations exactly like standard cold dark matter
particles and do not destroy all predictions of $\Lambda $CDM models.
However at smaller sub-galactic scales wormholes strongly interact with all
existing particles. They scatter photons, baryons, and dark matter particles
and, therefore, they do smooth away cusps predicted by numerical simulations
at galactic centers \citep{NFW}. Recall that cold heavy particles
unavoidably form cusps $\rho _{DM}\sim 1/r$, while observations %
\citep{G04,B03,W03} show not such a feature. This may be considered as an
essential indirect argument in favor of the existence of relic wormholes,
since all other known mechanisms of removing cusps are not efficient.

We point also out that strong theoretical arguments for the existence of
relic wormholes come from lattice quantum gravity \citep{AJL05}. Indeed, it
is assumed that at Planckian scales the topological structure of our
Universe should have fractal properties. During the inflationary stage the
topological structure of space should temper and may survive till the
present days in the form of relic cosmological wormholes. The problem of the
formation of relic wormholes is not described rigorously yet and we do not
discuss it here. Nevertheless, some hints on such a picture can be found
from the distribution of galaxies. On scales below $100Mp$ the distribution
of galaxies definitely shows fractal features \citep{L98}, see also the more
resent results in \citep{CIR14}. Such a structure may serve as a direct
trace of the actual topological structure of space. Indeed, if we assume the
homogeneous distribution of galaxies then the number counts $N(R)\sim R^D$
(where $N(R)$ is the number of galaxies within the radius $R$ and $D$ is the
dimension) reflects the behavior of the physical volume of space. It crosses
over to the homogeneity only on larger scales \citep{Planck18,Planck24}
which however cannot rule out the possibility of the existence of relic
wormholes. Indeed, as it was shown by \citep{B16} in the absence of peculiar
velocities wormholes do not perturb spectrum of the cosmic background
radiation and, therefore, they cannot be distinguished on the sky. The
detection of relic wormholes requires studying more subtle effects. Of the
primary interest are those effects which can be disentangled from effects
produced by black holes and other forms of matter.

In the present paper we consider the scattering of background radiation on
wormholes and show that they can be in principle observed by means of the
effect analogous to the kinematic Sunyaev-Zel'dovich effect (KSZ) %
\citep{ZS,ZSa}. KSZ signal is based on the inverse Compton scattering of
relic photons on a moving electron gas. It represents one of the main tools
in studying peculiar motions of clusters and groups of galaxies, e.g., see %
\citep{KSZ1,KSZ2,KSZ3}, and see also more applications in a recent review %
\citep{B16}. It is actually produced by any kind of matter which scatters
CMB (not only by a hot electron gas). As it was shown in \citep{B16}, in the
first order in $V/c$ the contribution of wormholes into KSZ cannot be
separated from that of the electron gas in clusters and groups. Therefore,
there are two possibilities. First one is to look for such an effect in
those spots on the sky where the baryonic matter is absent, e.g., in voids
where the leading contribution will come from wormholes alone. In this case
however we need also some additional independent effects to be sure that the
signal comes from a void and not from the last scattering sphere. The second
possibility is to study next order corrections and peculiar features of the
scattering of background radiation on wormholes. In the case of CMB it turns
out that already in the second order KSZ effect on wormholes differs from
that on other sorts of matter. In the case of high-energy cosmic rays KSZ
produces simply a shift of spectrum without the change of it's form.

The most simple wormhole is described by a spherically symmetric
configuration. Spherical wormholes can be made stable only by the presence
of exotic matter \citep{HVis98}. While the natural sources of the exotic
matter are not found, we should state that all relic spherical wormholes
collapse very rapidly and hardly survive till the present days. If this
occur at relatively late time compared to the time of photon decoupling,
then emissions from collapsed structures may contribute to the cosmic-ray
background which is different from CMB (e.g., infrared, X-ray, etc.).
Remnants of such spherical wormholes can not be distinguished from ordinary
primordial black holes and we do not discuss them here. However, as it was
shown recently by \citep{ks16} stable relic wormholes may exist without
exotic matter, if they have a less symmetric structure. The rate of
evolution of such wormholes is comparable with the rate of cosmological
expansion and, therefore, such wormholes may survive till the present days.
The less symmetric wormholes have throat sections in the form of a torus or
even more complicates surfaces \citep{ks16}. We use a torus-like wormhole in
considering some peculiar features of the scattering on a single wormhole is
the section 4. Torus-like wormholes have random orientations in space. Upon
averaging over orientations the torus-like wormhole acquires features of a
spherically symmetric configuration. This allows us to use spherical
wormholes in considering estimates for KSZ and the second order corrections
to KSZ.

\section{Cross-sections and KSZ effect}

The scattering of electromagnetic waves on a spherical wormhole has been
considered first by \citep{sct,sct2}. There are two important features of
such a scattering which are the generation of a specific interference
picture upon scattering on a single wormhole \citep{KSWS} and the generation
of a diffuse halo around any discrete source \citep{KSS}. If a wormhole is
not very big, the interference picture gives too weak signal and, therefore,
it can be used only in the future observations. The generation of the
diffuse halo around discrete sources may have various interpretations and
this results in a difficulty to disentangle effects of wormholes and, for
example, effects of the scattering on dust.

The simplest model of a spherical wormhole is given by a couple of
conjugated spherical mirrors, when a relict photon falls on one mirror it is
emitted, upon the scattering, from the second (conjugated) mirror. Such
mirrors represent two different entrances into the wormhole throat and they
can be separated by an arbitrary big distance in the outer space. The
cross-section of such a process has been described by \citep{KSWS} and can
be summarized as follows. Let an incident plane wave (a set of photons)
falls on one throat. Then the scattered signal has two components. The first
component represents the standard diffraction (which corresponds to the
absorption of CMB photons on the throat) and forms a very narrow beam along
the direction of the propagation. This is described by the cross-section
\begin{equation}
\frac{d\sigma _{absor}}{d\Omega }=\sigma _{0}\frac{\left( ka\right) ^{2}}{%
4\pi }\left\vert \frac{2J_{1}\left( ka\sin \chi \right) }{ka\sin \chi }%
\right\vert ^{2},  \label{abs}
\end{equation}%
where $\sigma _{0}=\pi a^{2}$, $a$ is the radius of the throat, $k$ is the
wave vector, and $\chi $ is the angle from the direction of propagation of
the incident photons, and $J_{1}$ is the Bessel function. Together with this
part the second throat emits an omnidirectional isotropic flux with the
cross-section
\begin{equation}
\frac{d\sigma _{emit}}{d\Omega }=\sigma _{0}\frac{1}{4\pi }.  \label{flux}
\end{equation}%
The both total cross-sections coincide 
\begin{equation*}
\int \frac{d\sigma _{absor}}{d\Omega }d\Omega =\int \frac{d\sigma _{emit}}{%
d\Omega }d\Omega =\sigma _{0}
\end{equation*}
which expresses the conservation law for the number of absorbed and emitted
photons. In the absence of peculiar motions (a static gas of wormholes)
every wormhole throat end absorbs photons as the absolutely black body,
while the second end re-radiates them in an isotropic manner (\ref{flux})
with the same black body spectrum. It is clear that there will not appear
any distortion of the CMB spectrum at all. In the presence of peculiar
motions the motion of one end of the wormhole throat with respect to CMB
causes the angle dependence of the incident radiation with the temperature
$$
T_{1}=\frac{T_{CMB}}{\sqrt{1-\beta _{1}^{2}}\left( 1+\beta _{1}\cos \theta
_{1}\right) }
\simeq
$$
$$
 T_{CMB}\left( 1-\beta _{1}\cos \theta _{1}+\frac{1}{2}%
\left( 1+2\cos ^{2}\theta _{1}\right) \beta _{1}^{2}+...\right)
$$
where $\beta _{1}=V_{1}/c$ is the velocity of the throat end and $\beta
_{1}\cos \theta _{1}=\left( \vec{\beta}_{1}\vec{n}\right) $, $\vec{n}$ is
the direction for incident photons. Therefore, the absorbed radiation has
the spectrum%
$$
\rho \left( T_{1}\right) =\rho \left( T_{CMB}\right) +\frac{d\rho \left(
T_{CMB}\right) }{dT}\Delta T_{1}+
$$
$$
+\frac{1}{2}\frac{d^{2}\rho \left(
T_{CMB}\right) }{dT^{2}}\Delta T_{1}^{2}+...,
$$
where $\rho \left( T\right) $ is the Planckian spectrum and $\Delta
T_{1}=T_{1}-T_{CMB}$. As it was discussed previously by \citep{B16} in the
first order in $\beta _{1}$ the above anisotropy does not contribute to the
re-radiation of relic photons from the second end according to (\ref{flux}).
Indeed, integration over the incident angle $\theta _{1}$ gives $%
\left\langle \Delta T\right\rangle =-\frac{1}{4\pi }\int $ $\beta _{1}\cos
\theta _{1}d\Omega =0$. In other words, in the firs order in $\beta _{1}$
the peculiar motions of the absorbing ends of wormholes can be ignored. In
this case the KSZ effects caused by wormholes and by the standard baryonic
matter mix and cannot be disentangled. The difference however appears in the
second order in $\beta _{1}$. Indeed, considering the second order we find
$$
\left( \Delta \rho \right) _{2}=T_{CMB}\frac{d\rho \left( T_{CMB}\right) }{dT%
}\frac{\left( 1+2\left\langle \cos ^{2}\theta _{1}\right\rangle \right) }{2}%
\beta _{1}^{2}+
$$
$$
+\frac{T_{CMB}^{2}}{2}\frac{d^{2}\rho \left( T_{CMB}\right) }{%
dT^{2}}\left\langle \cos ^{2}\theta _{1}\right\rangle \beta _{1}^{2}
$$
where $( \Delta \rho ) _{2}=\left\langle \rho \left( T_{1}\right) -\rho
\left( T_{CMB}\right) \right\rangle _{2}$ which gives
$$
\left( \Delta \rho \right) _{2}=\frac{\beta _{1}^{2}}{6T_{CMB}^{3}}\frac{d}{%
dT}\left( \frac{d\rho \left( T_{CMB}\right) }{dT}T_{CMB}^{5}\right) ,
$$
where we used $\left\langle \cos ^{2}\theta _{1}\right\rangle =\frac{1}{3}$.
This means that together with the standard Planckian spectrum $%
I(T_{CMB})=c\rho (x) =I_{0}\frac{x^{3}}{e^{x}-1}$, where $I_{0}=\frac{2h}{%
c^{2}}\left( \frac{k_{B}T_{CMB}}{h}\right) ^{3}$ and $x=h\nu /k_{B}T_{CMB}$,
every wormhole emits the additional isotropic flux of photons with the
spectrum%
\begin{equation}
\left( \frac{\Delta I}{I_{0}}\right) _{2}=\beta
_{1}^{2}=f(x)\beta _{1}^{2},  \label{sp}
\end{equation}%
where $f(x)=\frac{1}{6}\frac{x^{4}e^{x}\left(
3e^{x}-3+x\left( e^{x}+1\right) \right) }{\left( e^{x}-1\right) ^{3}}$.
We point out that in this case the distortion of the spectrum does not
reduce to a frequency-invariant shift of the temperature. For the sake of
comparison we plot the function $f(x)$ (dotted line for $\beta _{1}^{2}=0.05$%
), the standard Planckian spectrum (in circles) and the sum (solid line) on
Fig.1. 
\begin{figure}[tbp]
\plotone{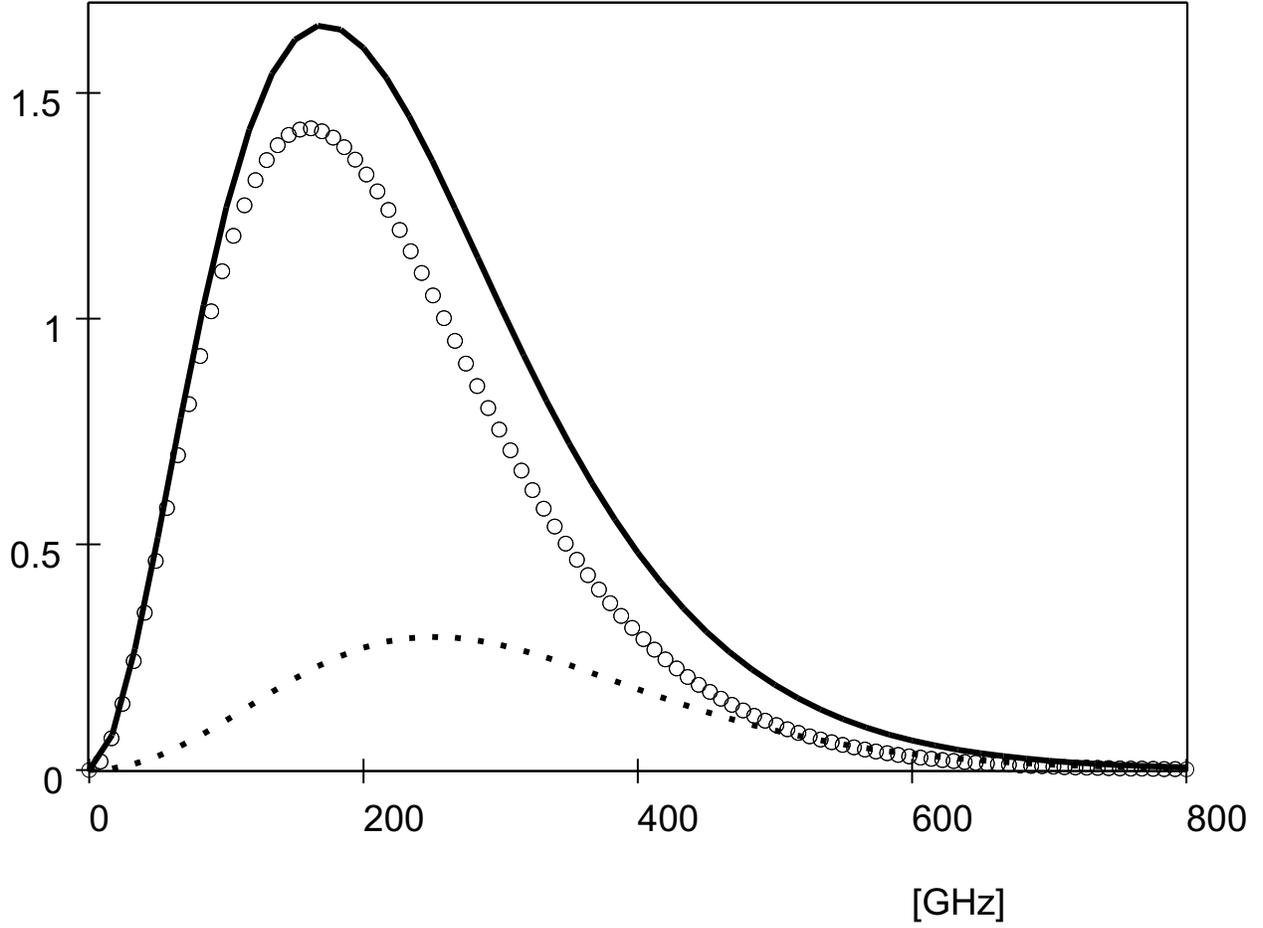}
\caption{Spectra emitted by a single wormhole. In circles is the standard
CMB, dotted line is the additional flux (\protect\ref{sp}) for $ \protect%
\beta _{1}^{2}=0.05$, and the solid line is the sum. }
\label{fig1}
\end{figure}
The estimate of the relative integrated amplitudes of the radiation
emitted by a single wormhole is given by
\begin{equation}
\frac{\left( \Delta I\right) _{2}}{I_{CMB}}=\allowbreak 5.\,\allowbreak
33\times \beta _{1}^{2}.  \label{2ksz}
\end{equation}

When considering a cloud of wormhole throats, in addition to the standard
CMB every throat radiates photons with the flux $(\Delta I)_{2}$. In the
presence of peculiar velocities the CMB part undergoes the Doppler shift
(which is the complete analog of the KSZ effect) 
$$
\frac{\Delta T_{KSZ}}{T_{CMB}}=\beta _{p}\tau _{w}.
$$
Here $\beta _{p}$ is the projection of the peculiar velocity of the cloud
along the line of sight and the optical depth $\tau _{w}$ defined as
\begin{equation}
\tau _{w}=\int \pi a^{2}n(r,a)dad\ell ,  \label{tau}
\end{equation}%
where the integration is taken along the line of sight and $n(r,a)$ is the
number density of wormholes measured from the center of the cloud and
depending on the throat radius $a$. The optical depth $\tau _{w}$ is
interpreted as follows. Let $L$ be the characteristic size of the cloud of
wormholes. Then on the sky it will cover the surface $S\sim L^{2}$, while
the portion of this surface covered by wormhole throats is given by
$$
\tau _{w}=\frac{N\pi \overline{a^{2}}}{L^{2}}=\pi \overline{a^{2}}\overline{n%
}L,
$$
where $N$ is the number of wormhole throats in the cloud and $\overline{n}$
is the mean density. In a sufficiently dense cloud $\tau _{w}\sim 1$ this
effect produces simply a hot or a cold (depending on the sign of $\beta _{p}$%
) spot on the CMB maps. It is important that KSZ corresponds to a
frequency-invariant temperature shift which leaves the primary CMB spectrum
unchanged.

The second order effect discussed earlier does not depend on velocities of
throats in the cloud. It however depends on velocities of conjugated
entrances into throats and is given by%
\begin{equation}
\frac{\left( \Delta I\right) _{2KSZ}}{I_{CMB}}=\allowbreak 5.
33\times \left\langle \beta _{1}^{2}\right\rangle \tau _{w},  \label{rt}
\end{equation}%
where $\left\langle \beta _{1}^{2}\right\rangle \tau _{w}=\int \beta _{1}^{2}\pi a^{2}n(r,a,\beta _{1}^{2})dad\ell d\beta
_{1}$.
In general, such an effect is very small, since the typical values does not
exceed $\left\langle \beta _{1}^{2}\right\rangle \sim 10^{-4}$. It is
however measurable for sufficiently dense clouds $\tau _{w}\sim 1$ and which
is important it cannot be reduced to a shift of CMB temperature and,
therefore, it does slightly change the primary CMB spectrum according to (%
\ref{sp}). This gives a new tool which allows to distinguish the
contribution of wormholes into KSZ effect from that of the rest matter.

\section{Cosmic-ray spectrum and KSZ effect}

Measurements of the High-energy cosmic-ray spectrum of electrons and
positrons is described by a smoothly broken power- law model, e.g., see %
\citep{RS}%
\begin{equation}
\Phi(E)=\Phi_0\left(\frac{E_0}{E}\right)^{\gamma _1 }\left[1+\left(\frac{E}{%
bE_0}\right)^{\frac{\gamma _2-\gamma _1}{\Delta}}\right]^{-\Delta},
\label{pl}
\end{equation}%
where $\Delta=0.1$, $\Phi_0=A/E_0$, $E_0=100$GeV, $A=(1.64 \pm 0.01)\times
10^{-2}$ $m^{-2} s^{-1} sr^{-1} $, $b=9.14 \pm 0.98$, $\gamma _1 = 3.09 \pm
0.01$, and $\gamma _2 =3.92 \pm 0.20$. It shows that at energies $E\simeq
E_b=bE_0$ the spectral index changes from $\gamma _1 \approx 3.1$ to $\gamma
_2\approx 3.9$ \citep{RS}. The cross-section described in the previous
section works in the case of rays as well. First we point out that the
presence of relic wormholes leads to the formation of a diffuse halo (of a
low intensity) around any discrete source. When wormholes do not move in
space, then they do not change the spectrum at all \citep{KSZ08}. Consider
now an incident on a wormhole particle. In the rest frame of the wormhole
the energy of the incident particle changes according to the standard
Lorentz transformation. For High-energy particles it gives%
\begin{equation}
E^{\prime }=\frac{E+\left( Vp\right) }{\sqrt{1-\beta ^{2}}}\simeq E\left(
1+\beta \frac{cp}{E}\cos \theta \right) \simeq
\end{equation}
$$
 \simeq E\left( 1+\beta \cos \theta
\right) .
$$
Here $\cos \theta $ is the angle between the direction of the incident
particle and the velocity of a wormhole entrance. Thus the change of the
energy of the particle is given by%
\begin{equation}
\frac{\Delta E^{\prime }}{E}=\frac{1}{\sqrt{1-\beta ^{2}}}\left( 1+\beta
\sqrt{1-\frac{m^{2}c^{4}}{E^{2}}}\cos \theta \right) -1
\end{equation}%
$$
\frac{\Delta E^{\prime }}{E}\simeq \beta \cos
\theta .
$$
According to (\ref{flux}) the incident particles produce the isotropic flux
from the second entrance into the wormhole throat (in the rest frame of the
second entrance) with the same energy $E^{\prime }$. In the case of an
isotropic distribution of incident particles the mean change of the energy
vanishes $\left\langle \cos \theta \right\rangle =0$. This means that for
the isotropic background we have the same situation as in the case of CMB,
the motion of the absorbing end of the wormhole does not matter. The
peculiar motion of the emitting throat entrance produces the effect
analogous to KSZ effect, which is the common Doppler shift of the energy $%
\Delta E^{\prime }/E=\beta _{p}$, where $\beta _{p}=V_{p}/c$ is the
projection of the wormhole velocity on the direction pointing out to the
observer. For the spectrum (\ref{pl}) it can be described in terms of the
respective shift of the value $\Delta E_0/E_0=\beta _{p}$ which admits both
signs. The basic property of the spectrum (\ref{pl}) is that such a shift
does not change the form of the spectrum. In the case $\tau _{w}\ll 1$ the
form of the spectrum does not change also in next orders in $\beta _{p}$.
For sufficiently dense clouds of wormholes $\tau _{w}\sim 1$ relativistic
corrections include also terms of the form $\Delta E^{\prime }/E\sim
m^2c^4/E^2$ which do produce distortions of the form of the spectrum but
they are to small for high energies and can be neglected. The Doppler shift
appears also in the case when the source of radiation moves and both effects
merge. KSZ however somewhat smoothes the break in the spectral index at the
energy $E\simeq E_b\simeq 0.9$TeV. Thus the basic effect of relic wormholes
which admits observation is a small shift (positive or negative) of the
value $E_b$ in high-energy cosmic rays.

In conclusion of the section we point out that such a mechanism (the
generation of a shift of spectrum) works during the whole period of the
evolution of the Universe. In particular, it works also at the time of
photon decoupling and if there were such processes as dark-matter particle
annihilation or decay, effects of scattering on wormholes should be
imprinted in the spectrum.

\section{The scattering of CMB on a single torus-like wormhole}

In the case of a single wormhole we should account for the two important
features. The first feature is the fact that a stable cosmological wormhole
has the throat section in the form of a torus \citep{ks16}. The simplest
model of a torus-like wormhole is given by a couple of conjugated torus-like
mirrors. Therefore, if such a wormhole is sufficiently big, then the
simplest way to find it is to look for the direct imprints on CMB maps.
Indeed, by means of KSZ effect a wormhole should produce a ring on CMB map
that has a temperature which is slightly different from the background
temperature. In particular, it was reported recently in \citep{MNR}, that
there are, with confidence level 99.7 per cent, such ring-type structures in
the observed cosmic microwave background. We hope that such structures could
be imprints of cosmological wormholes indeed. In this case however more
frequent structures should have elliptical form, since tori (wormhole
throats) have random orientations in space.

The second important feature is that the scattering forward (i.e. absorption
of CMB photons (\ref{abs})) produces much bigger effect (since $kR\gg 1$, $%
ka\gg 1$, where $k$ is the wave-vector, $R$ is the largest, and $a$ is the
smallest radiuses of the torus respectively). This effect corresponds to the
standard diffraction on the torus-like obstacle. In the approximation $\mu
=a/R\ll 1$, where $a$ is the smallest radius of the torus, we may use the
flat screen approximation.

Let the orientation of the torus (the normal to the torus direction) be
along the $Oz$ axis, i.e. $m=(0,0,1)$. The cross-section depends on the two
groups of angle variables, i.e. the two unit vectors $n_{0}(\phi _{0},\theta
_{0})$ and $n(\phi ,\theta )$. The vector $n_{0}=(\cos \phi _{0}\sin \theta
_{0},\sin \phi _{0}\sin \theta _{0},\cos \theta _{0}$) points to the
direction of the incident photon (i.e., the wave vector is $k_{0}=\frac{%
\omega }{c}n_{0}$), while the vector $n$ corresponds to the scattered
photons. Then the cross-section is given by 
\begin{equation*}
\frac{d\sigma }{d\Omega }=\sigma _{R}\sin ^{2}\theta _{0}\frac{\left(
kR\right) ^{2}}{4\pi }\left( \frac{1+\cos ^{2}\theta }{2}\right) \left\vert
F\right\vert ^{2},
\end{equation*}
where $\sigma _{R}=\pi R^{2}$, and the function $F$ is 
\begin{equation*}
F=\left( 1+\mu \right) ^{2}\frac{2J_{1}\left( \left( 1+\mu \right) y\right)
}{\left( 1+\mu \right) y}-\left( 1-\mu \right) ^{2}\frac{2J_{1}\left( \left(
1-\mu \right) y\right) }{\left( 1-\mu \right) y}
\end{equation*}
where $y=kR\xi $. We also denote 
\begin{equation*}
\xi =\left( \sin ^{2}\theta +\sin ^{2}\theta _{0}-2\sin \theta \sin \theta
_{0}\cos \left( \phi -\phi _{0}\right) \right) ^{1/2}
\end{equation*}
and $J_{n}(y)$ are the Bessel functions. We also averaged $\sigma $ over
polarizations. Let us expand the kernel $F$ by the small parameter $\mu \ll
1 $ which gives%
\begin{equation*}
F\approx 2\mu \left( y\left( \frac{2J_{1}\left( y\right) }{y}\right)
^{\prime }+2\frac{2J_{1}\left( y\right) }{y}\right) .
\end{equation*}
Using the property $\left( J_{\nu }(y)/y^{\nu }\right) ^{\prime }=-J_{\nu
+1}(y)/y^{\nu }$ and the identity $J_{2}\left( y\right) =\frac{2}{y}%
J_{1}(y)-J_{0}(y)$ we get $F\approx 4\mu J_{0}(y) $ which gives
\begin{equation*}
\frac{d\sigma }{d\Omega }=8\sigma _{R}\frac{\left( ka\right) ^{2}}{4\pi }%
\left( 1-\cos ^{2}\theta _{0}\right) \left( 1+\cos ^{2}\theta \right)
\left\vert J_{0}(kR\xi )\right\vert ^{2} .
\end{equation*}
The intensity of the scattered radiation in the solid angle $d\Omega $ and
in the interval of frequencies $d\nu $ is given by
\begin{equation*}
\frac{1}{I_{\nu } }\frac{d\Delta I_{\nu }}{d\Omega }=\frac{2\sigma
_{R}\left( ka\right) ^{2}}{\pi }\left( 1+\cos ^{2}\theta \right) \int
\left\vert J_{0}(kR\xi )\right\vert ^{2}\sin ^{2}\theta _{0}d\Omega _{0},
\end{equation*}%
where $I_{\nu }=c\rho (x)$ is the intensity of the incident black body
radiation and $x=h\nu /k_{B}T_{CMB}$. The above expression improves the
absorbtion part given by (\ref{abs}). Since $kR\gg 1$, it shows the presence
of specific ring-type oscillations in the cross-section. Indeed, if we
consider the normal fall of the incident photons, i.e., $\theta _{0}=0$,
then we find $\frac{d\sigma }{d\Omega } \thicksim J_{0}(kR\theta)$. For
sufficiently remote throats the value $R\theta$ is small and such
oscillations should be imprinted in the diffraction picture of CMB in the
form of rings.

\section{Conclusion}

In conclusion we point out that in searching for KSZ signal from wormholes
we meet two basic problems. First one is the need of independent
observational effects related to wormholes which can be compared to KSZ. The
simplest effect of such a kind we find, if we consider propagation of cosmic
rays (of any origin) through the same region of space where we expect to
observe KSZ. Wormholes were shown to produce an additional damping in cosmic
rays \citep{KSZ08} which is determined by the same optical depth (\ref{tau})
$\tau _{w}$. Thus, if there is a discrete source of a standard intensity,
the optical depth can be directly measured. The damping is caused by the
capture of some part of particles by wormholes. Particles captured are
re-emitted (by the second entrance into wormhole throats) in an isotropic
way which forms a diffuse halo around any discrete source. In the absence of
peculiar motions of wormholes such a halo has the same energy spectrum.
Peculiar motions cause a shift of the initial cosmic ray spectrum without a
change of it's form. For example, random motions should somewhat smooth the
detected break in the teraelectronvolt cosmic-ray spectrum of electrons and
positrons \citep{RS}, while common peculiar motions simply produce an
additional shift of the threshold value $E_b$. The Doppler shift of the
spectrum can also be attributed to the motion of the source itself.
Therefore, to disentangle KSZ and the motion of the source represents very
difficult problem and such subtle effects require the further investigation.

Another possibility is to extract basic parameters (such as the density of
wormholes $n_{w}$ and the characteristic cross- section $\sigma _{0}=\pi
\overline{a^{2}}$) from the distribution of dark matter. For example, the
behavior of dark matter in galaxies may fix two parameters e.g., see %
\citep{PSS,KT} by means of measuring the empirical Newton's potential.
Indeed, in galaxies the distributions of dark and luminous matter strongly
correlate \citep{D04}. This means the existence of a rigid relation $\rho
_{DM}(k)=b(k)\rho _{vis}(k)$, where $\rho (k)$ are Fourier transforms for
dark and visible matter densities. Then from the observed distribution of
dark matter in galaxies, e.g. see \citep{G04,W03}, we may retrieve the
Newton's potential as \citep{KT}
\begin{equation*}
\phi _{emp}=-\frac{4\pi Gb(k)}{k^{2}} ,
\end{equation*}
which describes the deviation from the
Newton's law. Observations of rotation curves are fitted by the simple function $b(k)=\left( 1+(Rk)^{-\alpha }\right)$. At small scales ($Rk\gg 1$) it gives the standard Newton's
law, while at large scales $Rk\ll 1$ it transforms to the fractal law, or
the logarithmic behavior. We point out that the correction
observed in galaxies corresponds to the value $\alpha \approx 1$ and $R\sim
5Kpc$. These parameters can be related to the distribution of wormholes %
\citep{KS17} but this problem requires the further study.

The second problem is that in galaxies and clusters (as well in the hot
X-ray gas) the KSZ effect on CMB based on wormholes mixes with that on other
sorts of matter (dust, hot gas, etc.). The difference appears only in the
second order in $V/c$ (\ref{rt}) which requires sufficiently high accuracy
of observations.



\end{document}